# System Attack Modeling Techniques Critical Information Infrastructure

## A.K. Novokhrestov, A.A. Konev, A.S. Kovalenko, N.I. Sermavkin

Tomsk State University of Control Systems and Radioelectronics, 40 Lenina Ave., Tomsk, 634012 Russia
E-mail: kaa1@keva.tusur.ru

**Abstract.** Every day around the world, various organizations are exposed to more than a hundred attacks, most of which are successfully repelled by information security specialists. However, attacks are also carried out that some information systems or specialists are unable to repel, which is why a large number of enterprises, as well as individuals, suffer huge monetary and reputational losses. The aim of the work is to train specialists through cyber polygons and interactive games to a high level of knowledge and skills in the field of information security.

## 1. Introduction

It is quite obvious that to simulate attacks on CII systems, certain databases with the maximum possible number of currently known threats, vulnerabilities and approximate algorithms and techniques are needed, which together will represent a certain set of tools for designing attacks.

There are currently not many gadgets in this niche. If you ask ordinary information security specialists about the bases of threats or attacks known to them, then with a very high probability the majority will name the matrix MITER ATT & CK.

MITER divides ATT & CK into several different matrices: Enterprise (for corporate systems), Mobile (for mobile devices), and PRE-ATT & CK (for preparatory stages of attacks). Each of these matrices contains different tactics and techniques, due to the specifics of this matrix. The Enterprise Matrix consists of techniques and tactics that apply to Windows, Linux and / or MacOS systems. The Mobile Matrix contains tactics and techniques that apply to mobile devices.

The PRE-ATT & CK Matrix contains tactics and techniques for attackers to act before they attempt to exploit a vulnerability in a specific target network or system.

ATT & CK for Enterprise Matrix details the tactics and techniques that attackers use to infiltrate the network, compromise IT systems, escalate privileges, and navigate without detection. Early versions of the Matrix focused on corporate networks and local IT infrastructure. Over time, MITER has expanded the scope of ATT & CK for enterprises to include IaaS, PaaS and SaaS environments.

ATT & CK for Enterprise Matrix (v9) covers various desktop and server operating systems (Windows, macOS, Linux), cloud platforms (AWS, Microsoft Azure, Google Cloud Platform), SaaS solutions (Azure AD, Microsoft 365, Google Workspace) and network resources.

Key integrated tools and resources of MITER ATT & CK:

- ATT & CK Navigator (a tool for matching controls with ATT & CK technicians);
- Malware Archeology Windows ATT & CK Logging Cheat Sheet (a number of Windows event logging cheat sheets).
- Uber Metta (an open source Uber project that mimics the actions of intruders and is coordinated with MITER ATT & CK).
- MITER Cyber Analytics Repository (a reference site with various analytics that can be used to identify the behavior specified in MITER ATT & CK).
- MITER Caldera (an open source automated intruder simulation tool based on MITER ATT & CK).
- ATT & CK Tableau Table by Cyb3rPanda (ATT & CK in a public instance of the interactive business intelligence system).
- Palo Alto Unit 42 Playbook Viewer (a free Playbook Viewer that shows the known behaviors of several groups of attackers, correlating them with MITER ATT & CK data).
- Red Canary Atomic Red Team (open source tool from Red Canary to simulate attacker behavior with attachment to MITER ATT & CK).
- Anomali Cyber Watch (Free weekly report on key security and threat events of the week).
- Endgame Red Team Automation (Endgame Red Team Automation (an open source tool from Endgame that performs testing by simulating attacker behavior based on MITER ATT & CK data).

MITER has made a huge contribution to the development of the information security community by presenting the ATT & CK knowledge base and related tools, methods and approaches to ensuring security. Today, attackers increasingly find ways to act more and more stealthily and avoid detection using traditional security tools, specialists have to change their approach to detection and protection. ATT & CK shifts the focus away from low-level indicators such as IP addresses and domain names, encouraging us to look at attackers and defenses through a behavioral lens.

## 2. Formal approaches to attack modeling.

Typically, attack classification focuses on specific aspects of attack modeling. Such aspects can be formalisms for representing data and knowledge about attacks, levels of the OSI (Open Systems Interconnection) or

DoD (Department of Defense) models, types of attacks, scalability, the ability to take into account the dynamic characteristics of attacks (for example, time, parallel processes, chains of interconnected incidents), etc.

## 2.1 Modeling attack scenarios using discrete mathematics.

The importance of the problem of information security is growing every year, because cybercriminals develop and develop methods to gain access to confidential information, find new ways to violate the integrity and availability of data. In this regard, a lot of effort is spent on the development of defensive elements, which is why much less attention is paid to how to simulate an attack and how to present it.

When planning an attack, often, roughly speaking, the general features of the attack are first set, and after the formation of a superficial attack plan, this plan is deepened. So, if we consider the creation of a general concept of an attack as the first, initial level, then at the second level, the details of the steps that were determined at the first level will be carried out.

These levels can be specified formally. Let's say that each point of the attack plan at the first level will be a "word" that belongs to a formal language, specified by some formal grammar. A chain belongs to the attack script language generated by the grammar only if there is an derivation of it from the chain of this grammar. The process of constructing such an inference (and, consequently, determining whether a chain belongs to a language) is called parsing.

From a practical point of view, the most interesting is the analysis of context-free (CF) grammars. A context-free grammar is a special case of a formal grammar in which the left parts of all productions are single nonterminals (objects denoting some essence of the language (for example: a formula, arithmetic expression, command) and have no specific symbolic meaning).

For example a grammar like

$S \to aA \mid bB$

$A \to ab$

$B \to ba$

will be context free, since the left-hand sides of all productions are single nonterminals (S, A, B).

However grammars like

$S \to aB \mid bA$

$aA \to aa$

$bB \to bb$

will be context sensitive, since the left parts of all productions are not single nonterminals (aA, bB). The generative power of context-free grammars is sufficient to describe most of the syntactic structure of attacks. There are well-developed ways of solving the parsing problem for various subclasses of CS grammars.

For CS grammars, it is possible to introduce a convenient graphical representation of the output, called the output tree, and for all equivalent outputs the output trees are the same.

A tree is called an inference tree (or a parse tree) in the CS grammar $G_A = (V_N, V_T, S, P)$, if the following conditions are satisfied.

1. Each vertex of the tree is labeled with a symbol from the set $(V_N \cup V_T \cup \varepsilon)$, while the root of the tree is labeled with the symbol S; leaves are symbols from $(V_T \cup \varepsilon)$.

2. If a vertex of a tree is marked by a symbol $A \in V_N$, and its immediate descendants are marked by symbols $a_1, a_2, \ldots, a_n$, where each $a_i \in (V_T \cup V_N)$, then $A \to a_1 a_2 \ldots a_n$ is an inference rule in this grammar.

3. If a tree vertex is marked by the symbol $A \in V_N$, and its only immediate descendant is marked by the symbol $\varepsilon$, then $A \to \varepsilon$ is the inference rule in this grammar.

An example of an output tree is shown in Figure 1.

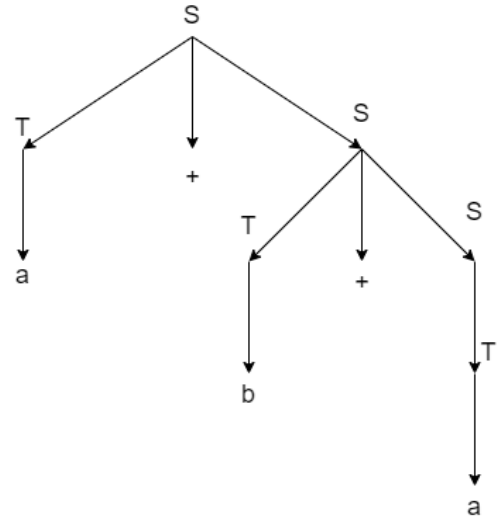

Fig. 1. An example of an output tree

The inference tree can be built in a top-down or bottom-up way.

In a top-down parsing, an output tree is formed from the root to the leaves; at each step up to the top marked with a nonterminal symbol, an attempt is made to find such an inference rule so that the terminal symbols contained in it are "projected" onto the symbols of the original chain.

The bottom-up parsing method consists in trying to "collapse" the original string to the initial character S; at each step, they look for a subchain that coincides with the right-hand side of some inference rule; if such a substring is found, then it is replaced by the nonterminal from the left side of this rule.

If the grammar is unambiguous, then for any construction method the same parse tree will be obtained

The attack scenario can also be represented in the form of a stochastic context-free grammar (CFG). This is a CFG, in which a probability corresponds to each inference rule.

For example a grammar like

0.6 NM $\to$ N FP

0.4 NM $\to$ N FP FP

will be a stochastic context-free grammar of 2 rules, since each rule is preceded by a probability (0.6 and 0.4).

Another example of CFG is shown in Figure 2 [1]. This model simplifies the modeling and presentation of attack scenarios.

$$\overset{0.25 \qquad 0.25 \qquad 0.25 \qquad 0.25}{s \to Aw_1U \mid Cw_1G \mid Gw_1C \mid Uw_1A}$$

$$\overset{0.1 \qquad 0.4 \qquad 0.4 \qquad 0.1}{w_1 \to Aw_2U \mid Cw_2G \mid Gw_2C \mid Uw_2A}$$

$$\overset{0.25 \qquad 0.25 \qquad 0.25 \qquad 0.25}{w_2 \to Aw_3U \mid Cw_3G \mid Gw_3C \mid Uw_3A}$$

$$\overset{0.8 \qquad 0.2}{w_3 \to GAAA \mid GCAA}$$

Fig. 2. An example of a stochastic context-free grammar

Another way to represent attack scenarios is using graphs, namely, graphs of computer attacks. An attack graph is a graph that provides all possible sequences of an attacker's actions to implement a threat. These se-quences of actions are called attack paths. An example of an attack graph is shown in Figure 3.

Another example of CFG is shown in Figure 2 [1].

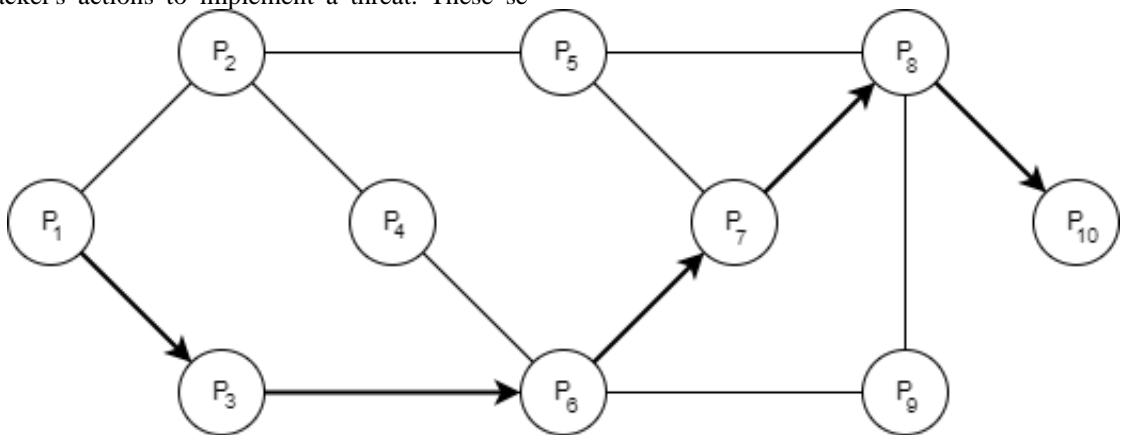

Fig. 3. An example of an attack graph

There are 3 types of attack graphs:

1) state enumeration graph - in such graphs the ver-tices correspond to triplets (s, d, a), where s is the source of the attack, d is the target of the attack, a is the ele-mentary attack; arcs indicate transitions from one state to another (Figure 4);

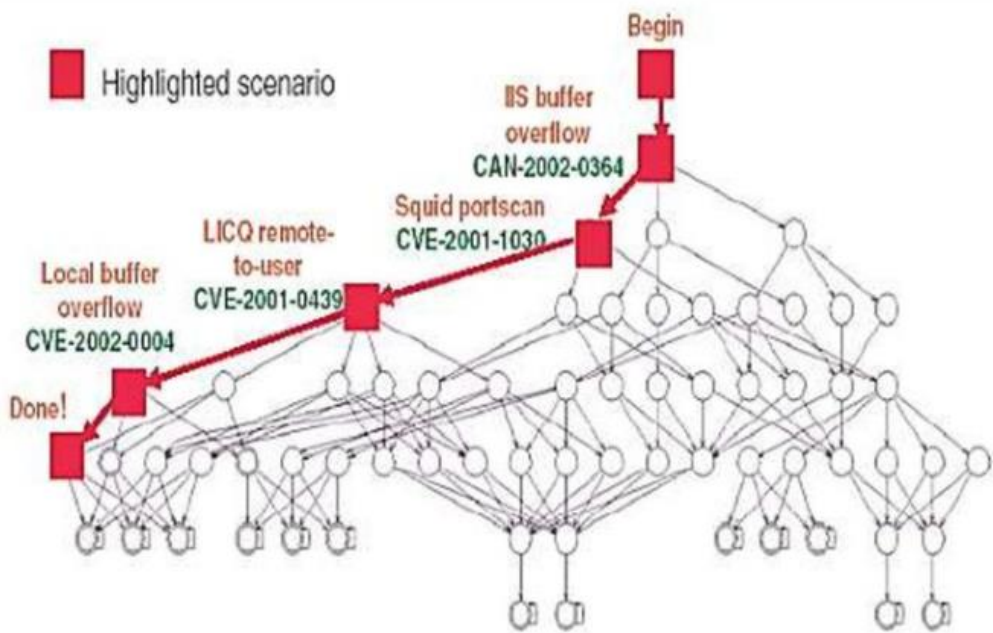

Fig. 4. State enumeration graph

2) condition-oriented dependency graph - the results of attacks correspond to vertices, and elementary attacks, leading to such results, correspond to arcs (Figure 5) [2];

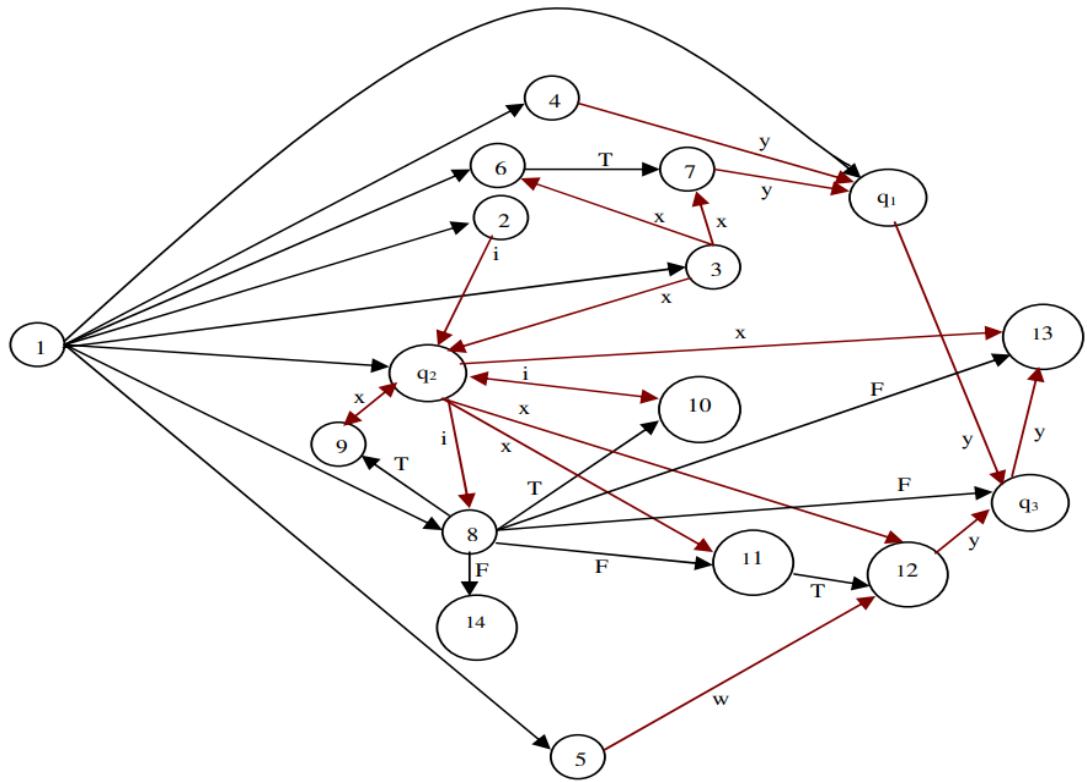

Fig. 5. Condition-oriented dependency graph

3) exploit dependency graph - vertices correspond to the result of attacks or elementary attacks, arcs display dependencies between the vertices - the conditions necessary to carry out the attack and the consequence of the attack (Figure 6) [3].

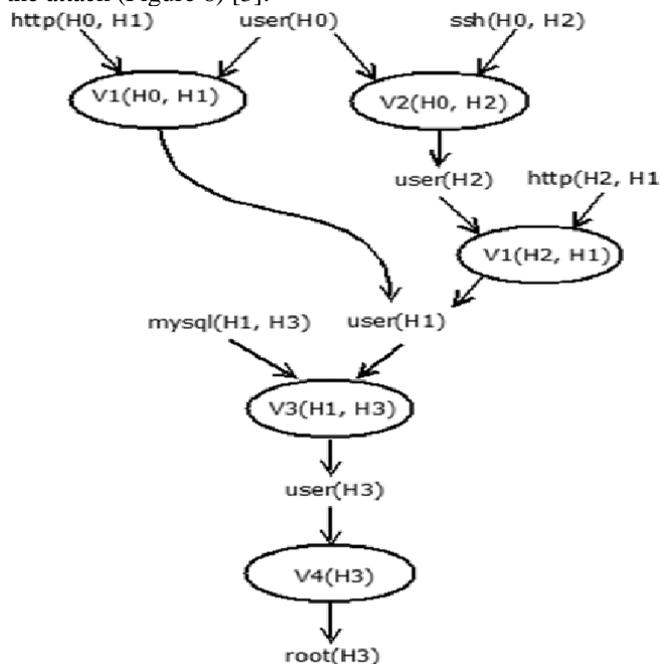

Fig. 6. Condition-oriented dependency graph

An atomic attack is understood as the use of a vulnerability by an intruder. An example of a rudimentary attack is, for example, an SSH service buffer overflow, allowing you to remotely gain system administrator rights.

The constructed graph contains all known attack scenarios for the attacker to reach threats. The result of its analysis may be:

1) a list of successful attacks not detected by IDS;

2) the ratio of the implemented security measures and the level of network security;

3) a list of the most critical vulnerabilities;

4) a list of measures to prevent the use of vulnerabilities in software for which there are no updates;

5) the smallest set of measures, the implementation of which will make the network secure.

The key problem of building an attack graph is scalability - the ability to build an attack graph for a network with a large number of hosts and vulnerabilities.

Another way of presenting attack scenarios is Petri nets. They are one of the most widely used formalisms for modeling attacks on computer networks. The most commonly used extensions to simulate attacks are colored Petri nets.

Petri nets have a number of parameters that make their use convenient for simulating attacks on a computer network. These include, in particular: the ability to graphically represent the model; convenience of model-ing dynamic and parallel processes; ability to reflect probabilistic processes; the ability to use time parameters; the presence of a large number of tools to support this formalism; ease of learning and use due to the presence of a small number of "primitives"; usability for analyzing various aspects of the security of a computer network. The lack of Petri nets, which is not compensated for by any of its extensions, is the inability to explicitly describe the behavior of the intruder and the attacked object, i.e. dynamic change of their states.

**2.2 Methods for representing (modeling) attack scenarios using simulation.**

With the help of simulation modeling, it is convenient to simulate the behavior of an intruder and an attacked object, for which specialists use the following four main approaches: discrete-event modeling, continuous dynamic modeling, system dynamics, and agent-assisted modeling. It should be noted that modern modeling systems are object-oriented, thereby ensuring the unification of all four approaches.

Let's consider some of the features of attack models implemented using agents. Typical examples of agents in large networks are viruses and worms. If it is required to simulate the behavior of the offender and the attacked object, then the use of artificial agents for this will reflect a whole set of behavioral characteristics, including activity, reactivity, autonomy, sociability, purposefulness, intentionality. This set of properties of agents makes them especially convenient for modeling distributed attacks, as well as cases of rivalry between violators. Most often, when simulating attacks using agents, multi-agent systems are built in which agents are divided according to the type of activity and combined into teams.

As an example, consider a multi-agent distributed denial of service (DDoS) attack model. To carry out this attack, the attacker must first compromise a large number of hosts and install attacking agents on them, which, in turn, will either simultaneously send a large number of network packets or difficult-to-process requests to the attacked hosts, or send too long or incorrect packets through intermediate nodes. ...

One of the options for building and using a multi-agent system in modeling DDoS attacks is the work where attack agents and protection agents are considered.

Attack agents are divided into the following two classes: "daemons", which directly implement the attack, and "masters", which coordinate the rest of the system's components.

Protection agents are divided into the following classes: information processing agents ("samplers"); attack detection agents ("detectors"); filtering and load balancing agents ("filters"); agents for identifying and incapacitating attack agents ("investigation agents").

In general, when modeling attacks using agents, in particular, such as attack and defense agents, general-purpose programming languages, for example, C ++, Java and others, as well as special languages for describing and implementing agents, for example, ACL, KQML and AgentTalk. A variety of such languages, as well as off-the-shelf libraries and agent development tools provide security professionals with ample opportunities to simulate attacks on large-scale networks.

**3. Meta Attack Language.**

As an example of a meta-attack language, enterpriseLang is considered for modeling threats to corporate systems. It is based on MAL (Meta Attack Language).

MAL is a threat modeling language framework that combines probabilistic attack and defense graphs with object-oriented modeling, which in turn can be used to create a digital subscriber line (DSL) and automate security analysis of model instances in each domain. It also provides a formalism that allows one to semi-automatically generate as well as efficiently compute very large attack graphs.

The MAL modeling hierarchy is shown in Figure 7.

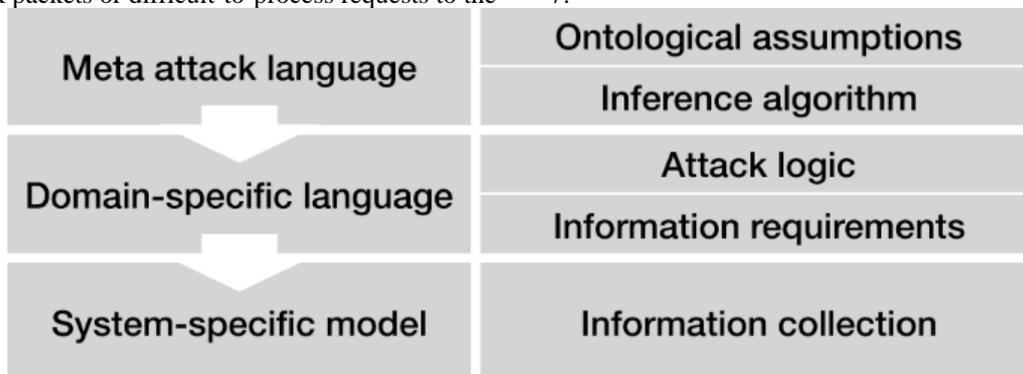

Fig. 7. MAL modeling hierarchy

The most common MAL syntax characters used by enterpriseLang to work with attack steps are related to each other and each of them is classified as "OR" (denoted by |) or "AND" (denoted by &). It is important to carefully analyze each step of the attack and find potential defenses as well as possible next steps in the attack. One successfully compromised attack step can lead to a second step. Sometimes, reaching the second step of an attack requires a combination of steps that have logical values indicating their status, where "enabled" or "disabled" is represented by setting the protection value to "TRUE" or "FALSE", respectively. If the value is FALSE, then the corresponding attack step can be reached, from which it is protected.

The enterpriseLang attack model creation process consists of three stages:

1. Extraction of information for each attacker's method from the ATT & CK Matrix;

2. converting the extracted information into MAL files;

3. unification of created files in one language.

An example of how the design of an attack on a corporate network using EnterpriseLang works is shown in Figure 8, where the MITER ATT & CK Matrix serves as input data for building the enterpriseLang threat modeling language, and enterpriseLang itself serves as input data for analyzing the behavior of attackers within the system model.

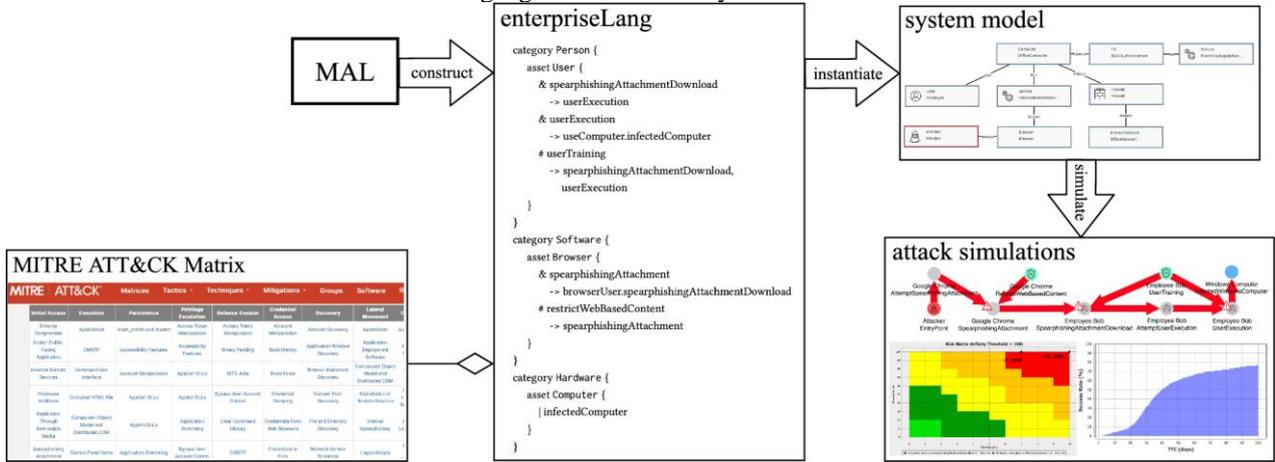

Fig. 8. Using EnterpriseLang for Enterprise Systems

By simulating attacks on a model of an enterprise system using the tools available, stakeholders can assess known threats to their enterprise, remedial actions, shortest attack paths that attackers can take in the simulated system, and the minimum time it takes for attackers to reach the various available stages of the attack and compromise the individual stages of the attack from the entry point.

After compiling the file, an attack graph is created as shown in Figure 9, where the circles represent the OR (|) attack steps, the squares represent the AND (&) attack steps, and the top side, the downward triangles, represent the defense.

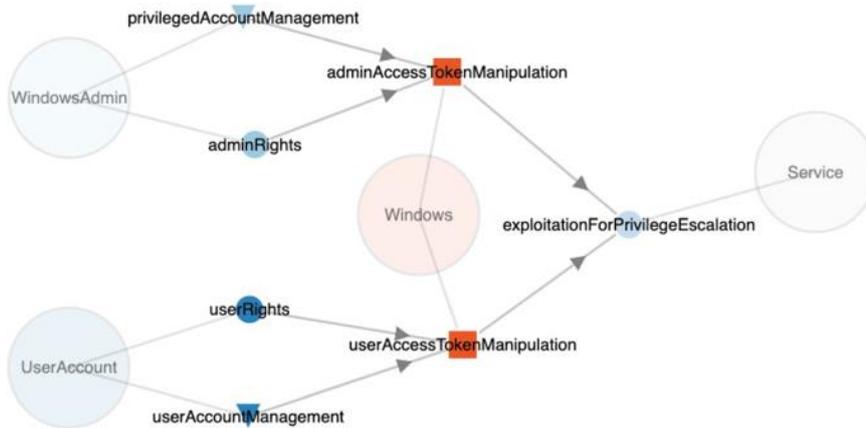

Fig. 9. Graphical representation of a simulated attack

The system model in the above example is quite small compared to real enterprise systems. System models created for real enterprise IT systems can be large, with thousands or millions of attack steps. Therefore, it is important to consider computational performance.

According to the MAL structure on which enterpriseLang is based, it is assumed that rational adversaries will take the shortest path to reach the attack step.

Therefore, the overall TTC (time to compromise) for the $A_{child}$ attack step (i.e. $T_{glob}(A_{child})$) is an estimate of the shortest time it takes opponents to reach any of their parent $A_{parent}$ plus the local time increment of this step of the attack (that is, $T_{local}(A_{child})$). In addition, the local TTC value of each attack step is selected from the assigned probability distributions [4].

For an "OR" attack step:

$$T_{glob}(A_{child}) = min(T_{glob}(A_{parent\ 1}), ..., T_{glob}(A_{parent\ n}) + T_{local}(A_{child})$$

For an "AND" attack step:

$$T_{glob}(A_{child}) = max(T_{glob}(A_{parent\ 1}), ..., T_{glob}(A_{parent\ n}) + T_{local}(A_{child})$$

The above algorithms are modified versions of the single-source shortest path (SSSP) algorithm [5], and the advantage of the modification is the ability to approximate attack steps while maintaining computational efficiency. In addition, the SSSP algorithm is deterministic. To perform probabilistic computations, a deterministic algorithm is used in Monte Carlo simulations. This creates a large set of plots with local TTC values for each attack step taken from their probability distributions. The SSSP algorithm is then used to compute the global TTC for each attack step in each attack graph. The resulting set of global TTC values for each attack step approximates the actual distribution.

On a 2016 Apple MacBook with Intel HD Graphics 515 and 1536MB of RAM, this algorithm was able to compute 1000 sample graphs with half a million nodes in less than three minutes. Thus, using relatively unimpressive hardware, surprisingly large networks can be computed [6].

Since the MAL, upon which the proposed EnterpriseLang is based, offers the possibility of using probability distributions to obtain more accurate results of attack modeling, probability distributions can be assigned to the steps of the EnterpriseLang attack and defense. These probability distributions are available from a variety of information sources [7], including vulnerability assessments and databases, previous research, vulnerability scanners, expertise, hacker knowledge, logs and alerts, and system health information. Therefore, by using the probability distributions assigned to the attack and defense steps instead of using binary relations, more accurate security estimates can be obtained, such as the probability of successful attack paths and risk matrices, as shown in Figure 10.

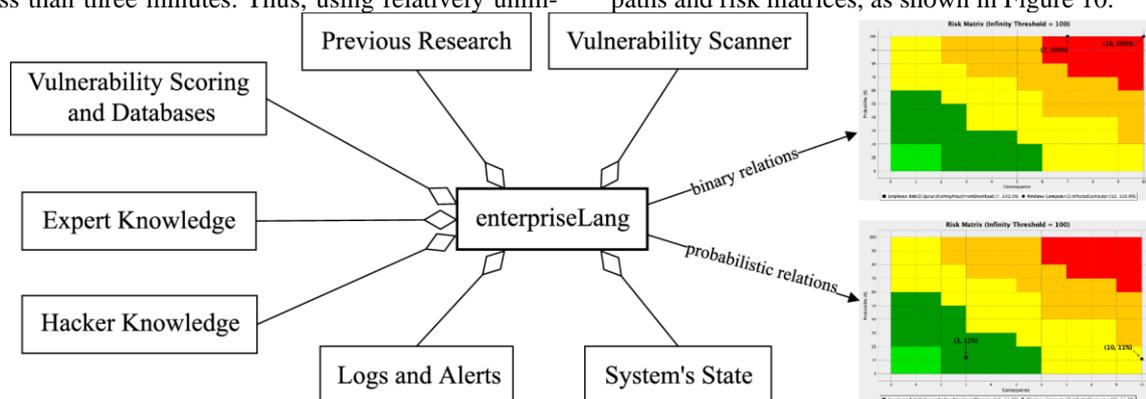

Fig. 10. Other sources that can be added to improve performance

The MAL enables security experts to codify domain-specific knowledge so that attacks on systems in the domain of interest can be simulated. The domain-specific attack modeling languages created in this way can then be used and reused by people with less security experience to automatically assess the security of specific systems in the domain.

## 4. Conclusion

In the course of the work, the knowledge bases of MITER ATT & CK were studied and considered in detail. Both the strengths and weaknesses of the NOS were identified for further productive work with them. It also considered the construction of attack models using them. Methods of presentation and modeling of attack scenarios using discrete mathematics were considered, the pros and cons of each method of presentation were identified.

## 5. Acknowledgments

This research was funded by the Ministry of Science and Higher Education of Russia, Government Order for 2020–2022, project no. FEWM-2020-0037 (TUSUR).